\documentclass[iop]{emulateapj}
\usepackage{graphicx}
\usepackage{amssymb}
\usepackage{epstopdf}
\usepackage{amsmath}
\usepackage{mathrsfs}
\usepackage{natbib}
\usepackage{setspace}
\usepackage{float}
\usepackage{times}
\defcitealias{Einasto2012a}{E12}
\defcitealias{Hwang2009}{HL09}
\defcitealias{Aguerri2010}{A10}

\slugcomment{Accepted for publication in The Astrophysical Journal}

\begin{document}

\shorttitle{Star Formation and Substructure in Clusters}
\shortauthors{Cohen et al.}

\title{Star Formation and Substructure in Galaxy Clusters}

\author{Seth A. Cohen, Ryan C. Hickox, Gary A. Wegner}
\affil{Department of Physics and Astronomy, Dartmouth College, 6127 Wilder Laboratory, Hanover, NH 03755, USA}

\author{Maret Einasto, Jaan Vennik}
\affil{Tartu Observatory, 61602 T\~{o}ravere, Estonia}

\begin{abstract}
We investigate the relationship between star formation (SF) and substructure in a sample of 107 nearby galaxy clusters using data from the Sloan Digital Sky Survey (SDSS).  Several past studies of individual galaxy clusters have suggested that cluster mergers enhance cluster SF, while others find no such relationship.  The SF fraction in multi-component clusters ($0.228 \pm 0.007$) is higher than that in single-component clusters ($0.175 \pm 0.016$) for galaxies with $M^{0.1}_{r} < -20.5$.  In both single- and multi-component clusters, the fraction of star-forming galaxies increases with clustercentric distance and decreases with local galaxy number density, and multi-component clusters show a higher SF fraction than single-component clusters at almost all clustercentric distances and local densities.  Comparing the SF fraction in individual clusters to several statistical measures of substructure, we find weak, but in most cases significant at greater than $2\sigma$, correlations between substructure and SF fraction.  These results could indicate that cluster mergers may cause weak but significant SF enhancement in clusters, or unrelaxed clusters exhibit slightly stronger SF due to their less evolved states relative to relaxed clusters.
\end{abstract}


\section{Introduction}
\label{sec:IntroSec}

A number of studies in recent years have measured the substructure of galaxy clusters \citep[e.g.,][]{Flin2006, Ramella2007, Einasto2010, Einasto2012b, Wen2013}.  These find that most clusters exhibit some amount of substructure in their spatial distributions \citep[e.g.,][hereafter E12]{Depropris2004, Einasto2012a}.  The presence of substructure and other deviations from symmetry are indicative of an active history of past or ongoing mergers of smaller groups and clusters \citep[e.g.,][]{Bird1993, Knebe2000}.  These mergers can affect the evolution of cluster galaxies through a variety of hydrodynamical and gravitational mechanisms that have been shown to both enhance and quench observed star formation \citep[SF; for a recent review, see][]{Boselli2006}.

For example, ram pressure caused by the interaction of the hot intragalactic medium with the cold interstellar medium has been shown to both increase cloud-cloud collisions and cloud collapse and promote cluster SF \citep[e.g.,][]{Evrard1991, Bekki2003, Bekki2010}, and sufficiently remove the ISM to stifle SF and subsequently turn these star-forming galaxies into post-starburst galaxies \citep[e.g.,][]{Bekki2003, Bekki2010}.  Galaxy-galaxy interactions can move gas from galaxy disks to circumnuclear regions, morphing spirals into lenticulars (and thus quenching SF).  Galaxy-cluster tidal forces, on the other hand, increase the kinetic pressure in the ISM and induce gas flow and increase SF \citep[e.g.,][]{Byrd1990, Elmegreen1997}.  Other mechanisms, such as viscous stripping, harrassment \citep[e.g.,][]{Moore1998, Mihos2004}, and starvation \citep[e.g.,][]{Bekki2002}, also compete to both ignite and cut off SF.

The relationship between SF and environment in clusters has been studied extensively with observations.  For example, the segregation of spiral and elliptical galaxies via density and cluster-centric radius is well-known \citep[e.g.,][]{Dressler1980, Depropris2004, Kodama2004, Rines2005}.  \citet{Lietzen2012} found that the fraction of elliptical galaxies increases, and the fraction of star-forming galaxies decreases, with both the richness of the host groups and the density of the surrounding large-scale supercluster environment.  A number of studies have recently addressed the issue of how cluster mergers disrupt this expected SF distribution by comparing cluster substructure and SF properties.  Several clusters show an enhanced presence of emission-line galaxies in clusters with substructure: for example, A1367 \citep{Cortese2004}, A3158 \citep{Johnston2008}, the ``bullet cluster" \citep{Rawle2010}, and A2465 \citep{Wegner2011}.  In the Group Environment and Evolution Collaboration (GEEC) group catalogue, \citet{Hou2012} found higher fractions of blue and star-forming galaxies in rich groups with substructure than in those without, but \citet{Hou2013} found no correlation between quiescent fraction and group dynamical state.  Similarly, \citet{Metevier2000} found high blue galaxy fractions in A98 and A115 that they attributed to the merging of subclusters, but they found no such correlation in A2356.  \citet{Tomita1996} found no enhancement of blue galaxies between the subclusters of A168, and \citet{Depropris2004} asserted that there is no dependence of blue galaxy fraction on probability of substructure, measured using the Lee-Fitchett statistics \citep{Fitchett1988}.  The work by \citet{Hwang2009} found enhanced SF in the region between the two subclusters of A168 and a lack of such enhancement in the same region in A1750.  Using the radial infall model of \citet{Beers1982}, they attributed these results to the differences in merger histories between the two clusters: A168's merger history indicated a merger had already occurred, while A1750 had not yet experienced a merger.

Other studies have estimated cluster merger history and determined the presence of so-called E+A galaxies \citep[e.g.,][]{Dressler1983, Dressler1999} to examine the relationship between cluster mergers and SF.  \citet{Ma2010} discovered that all E+A galaxies found in the major merger system MACS J0025.4-1225 lie between the dark matter peaks of the two clusters, suggesting that the merger $0.5 - 1$ Gyr ago both triggered SF and subsequently quenched it.  \citet{Shim2011} studied the mid-infrared (MIR) properties of the nearby merging cluster A2255 and found no evidence of enhanced star formation anywhere in the cluster.  Rather, the intermediate MIR-excess galaxies, representing galaxies transitioning from star-forming to passive states, suggested that cluster mergers tend to abruptly cease SF in member galaxies.  In the merging cluster A3921, \citet{Ferrari2005} found an abundance of actively star-forming galaxies between the two merging clusters and suggested that the merger could have triggered the SF.  However, they also found several post-starburst galaxies whose distribution and SF timescale could not be explained by the history of the cluster merger.

The discrepancy in these studies -- whether SF is enhanced, quenched, or unaffected by cluster mergers -- shows that the relationship between mergers and SF in the full cluster population remains poorly understood.  The past studies mentioned above focus mostly on individual systems, reporting local correlations between mergers and SF as they are discovered.  This has made it difficult to find any general trends between mergers and SF.  To address this problem, in this paper we compare the substructure and SF properties in the largest galaxy cluster sample of any such study to date.  Using spectroscopic data from the Sloan Digital Sky Survey Data Release 8 (SDSS DR8) and substructure information from \citetalias{Einasto2012a}, we compare the amount and distribution of substructure in 107 nearby galaxy clusters to observed star-forming properties by employing a variety of indications of substructure and robust determinations of SF.  In \S\ref{sec:DataSec} we describe our observational data set, and in \S\ref{sec:MethodsSec} we explain our methods of analysis.  We present our results from different comparisons between SF and substructure in \S\ref{sec:ResultsSec}, and discuss our interpretations in \S\ref{sec:DiscussionSec}.

Throughout our analysis we assume a standard cosmology of $H_{0} = 100\:h\:\textnormal{km}\:\textnormal{s}^{-1}\:\textnormal{Mpc}^{-1}$, $\Omega_{\textnormal{m}} = 0.27$, and $\Omega_{\Lambda} = 0.73$.

\section{Data}
\label{sec:DataSec}

Our galaxy cluster sample is taken from the catalogues of \citet{Tempel2012}, who employed the friends-of-friends method to identify galaxy groups and clusters in SDSS DR8 \citep{Aihara2011}.  From the 77,858 groups they identified, \citetalias{Einasto2012a} selected a subset of 109 objects that 1) have at least 50 members and are therefore robustly classified as clusters; and 2) have distances between $120 h^{-1}$ Mpc and $340 h^{-1}$ Mpc.  This upper limit was chosen as the distance above which the richness of groups decreases rapidly due to the effects of a flux-limited sample.  The lower limit was chosen to exclude nearby exceptionally rich clusters that must be analyzed separately (see \citealt{Tago2010} for details).  As explained below, we further remove clusters 68625 and 61613 \citepalias[cluster IDs from][]{Einasto2012a} from this subset to arrive at our final sample of 107 clusters.

\begin{figure}
\begin{center}
\includegraphics[scale=0.67]{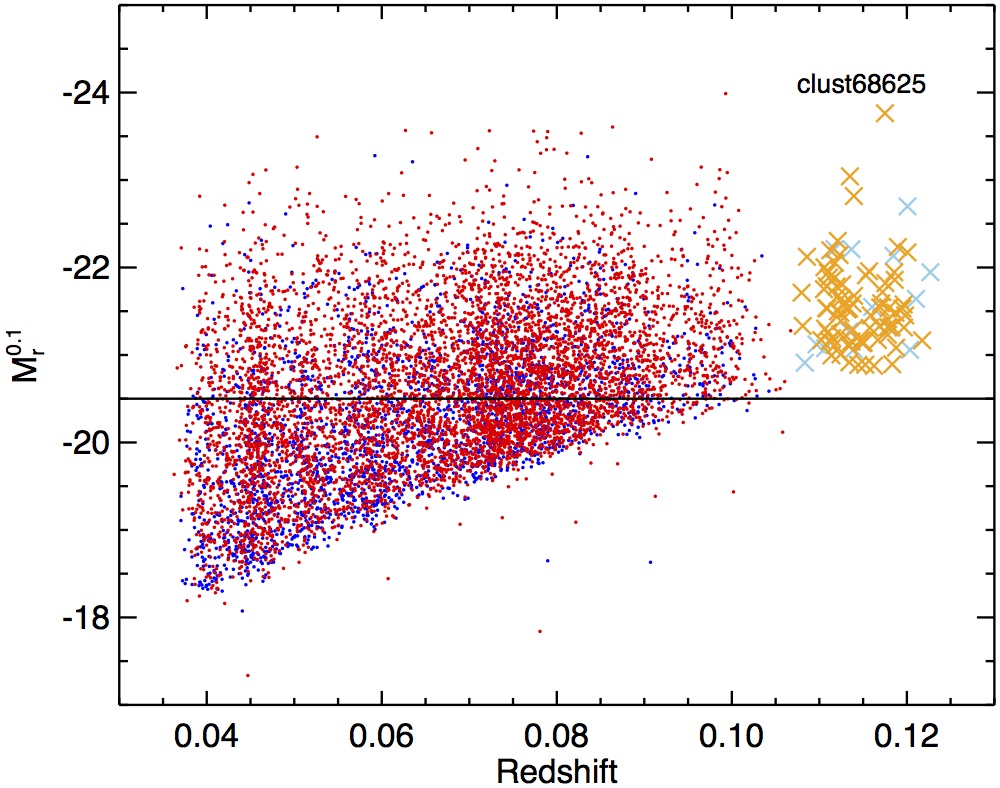}
\caption{\emph{r}-band absolute magnitude vs. redshift for all galaxies in the 109 clusters from \citetalias{Einasto2012a}.  The horizontal solid line denotes our cut in $M_{r}$.  Blue and light blue denote star-forming galaxies, while red and orange denote passive galaxies.  The light-colored X's show cluster 68625, which we remove due to its large distance and low completeness.}
\label{fig:MagCutFig}
\end{center}
\end{figure}

Much of our analysis relies on measurements of SF fraction in different regions of clusters, defined as the number of star-forming galaxies in a region divided by the total number of galaxies in that region.  As a consequence of a flux-limited sample, increasingly faint galaxies are unobserved in clusters at increasing redshift.  Since star-forming galaxies in clusters are preferentially fainter than passive galaxies, we must correct for this redshift effect.  To do this, we include in our analysis only those galaxies with $M^{0.1}_{r} < -20.5$, at which all of our clusters are complete.  As described in \citet{Hwang2009}, a galaxy's \emph{r}-band absolute magnitude is calculated from its apparent magnitude $m_{r}$ via 

\begin{equation}
\label{equ:MrEqu}
M^{0.1}_{r} = m_{r} - DM - K(z) - E(z),
\end{equation}
where $m_{r}$ is corrected for extinction; $DM \equiv 5\:\textnormal{log}(D_{L}/10\:\textnormal{pc})$ and $D_{L}$ is a luminosity distance; $K(z)$ is a \emph{K}-correction \citep{Blanton2007} to a redshift of 0.1, denoted by the superscript; and $E(z)$ is an evolution correction defined by $E(z) = 1.6(z - 0.1)$ \citep{Tegmark2004}.  Extinction-corrected magnitudes and \emph{K}-corrections are collected from the NYU Value-Added Galaxy Catalogue \citep[VAGC;][]{Blanton2005, Padmanabhan2008}.  Figure~\ref{fig:MagCutFig} plots $M^{0.1}_{r}$ as a function of redshift for the subset of 109 clusters identified by \citetalias{Einasto2012a}.  The horizontal solid line denotes our absolute magnitude cut.  Blue and light blue denote star-forming galaxies, while red and orange denote passive galaxies.  The light-colored X's illustrate our removal of cluster 68625, the only cluster whose distance is greater than $300 h^{-1}$ Mpc and whose $M^{0.1}_{r}$ limit would further constrain our magnitude cut.

We also check our results using an alternative method of correcting for redshift effects, which we explain briefly here.  We first determine the relationship between a cluster property (e.g., SF fraction) and redshift.  A linear fit to this trend produces a ``characteristic" measurement of this property at each redshift.  We then normalize the observed value of the property by its ``characteristic" measurement.  This effectively normalizes cluster properties by redshift while retaining intrinsic scatter in the data.  Furthermore, this method does not necessitate the sacrifice of information, as a luminosity cut warrants.  Our results remain consistent when implementing this normalization strategy instead of the absolute magnitude cut described above.  All analysis in this paper is thus performed using the absolute magnitude cut.

The SF properties of the cluster galaxies are measured using data from the Max Planck Institute (MPA)/Johns Hopkins University (JHU) VAGC \citep{Tremonti2004}.  This VAGC supplies SF information for almost all galaxies in the SDSS, including equivalent width (EW) and flux measurements for relevant emission and absorption lines, star formation rates (SFRs), stellar masses, and specific SFRs (sSFRs) for each galaxy.  These SFRs are determined using the techniques in \citet{Brinchmann2004} and references therein.  By cross-correlating the cluster members from \citetalias{Einasto2012a} with the galaxies from the VAGC, we have compiled SF information on all galaxies in our cluster sample.  We remove cluster 61613 from our sample, as only about 50\% of its galaxies are covered by the VAGC.

\begin{figure*}
\begin{center}
\includegraphics[scale=0.38]{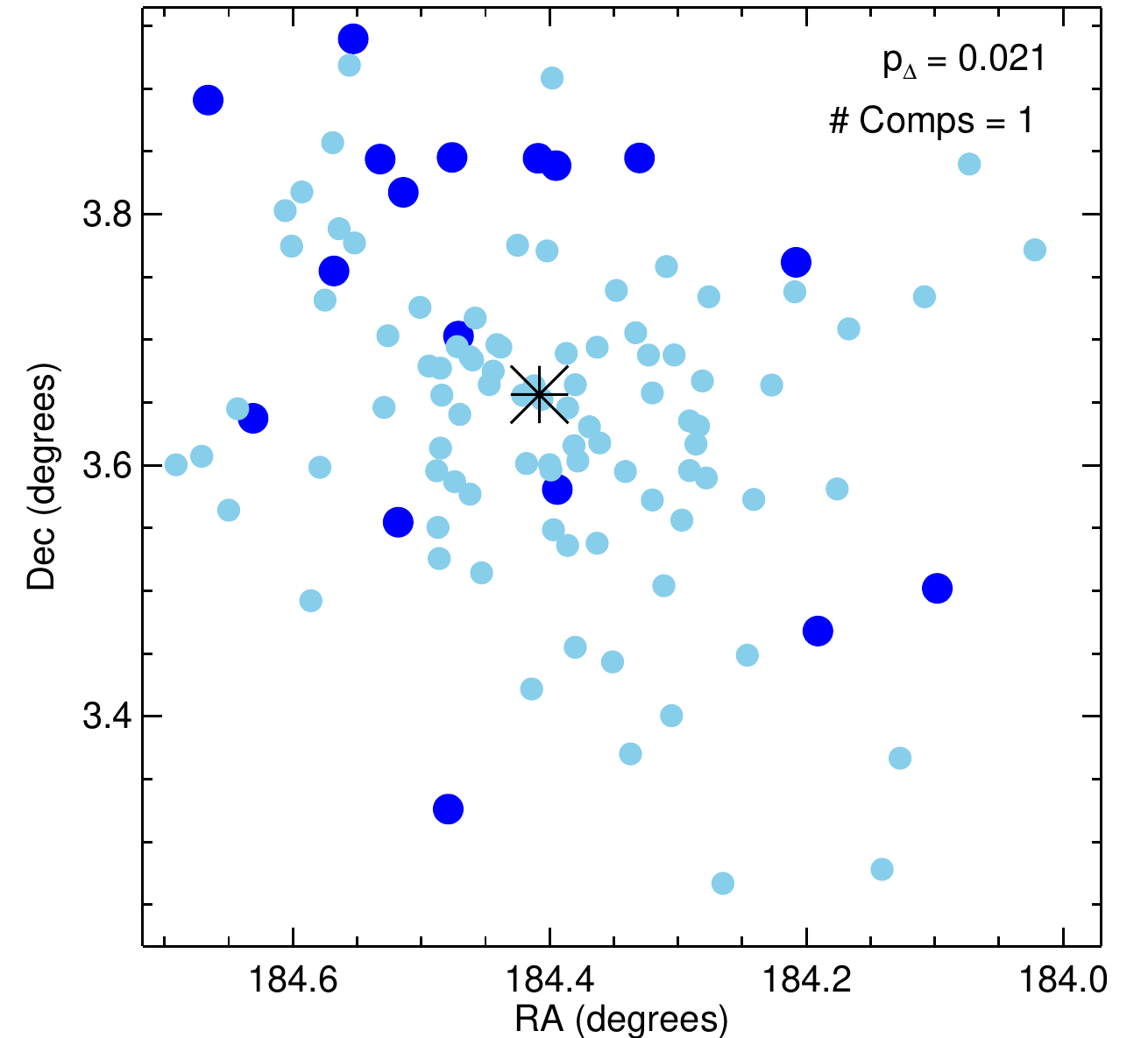}
\includegraphics[scale=0.38]{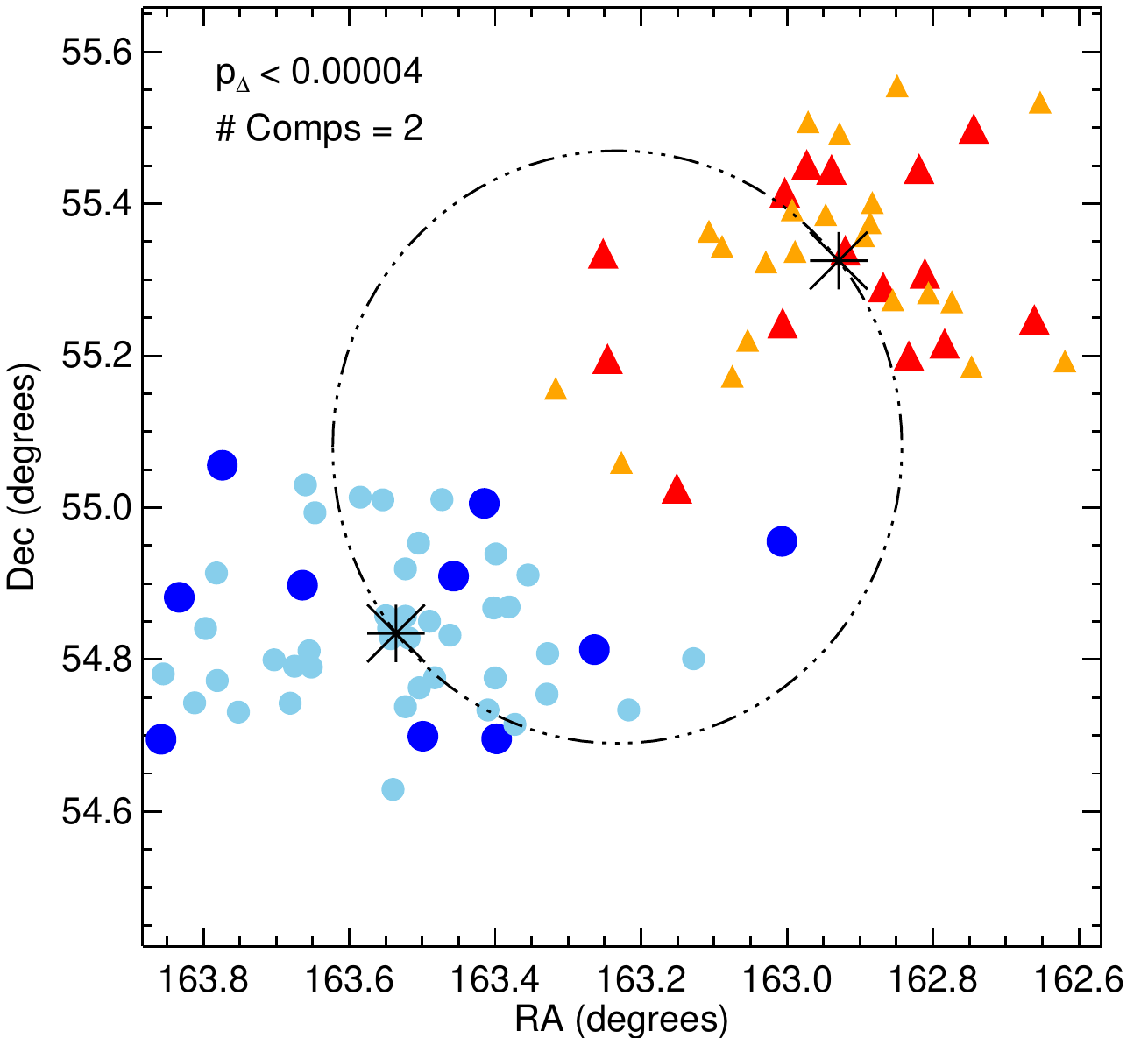}
\includegraphics[scale=0.38]{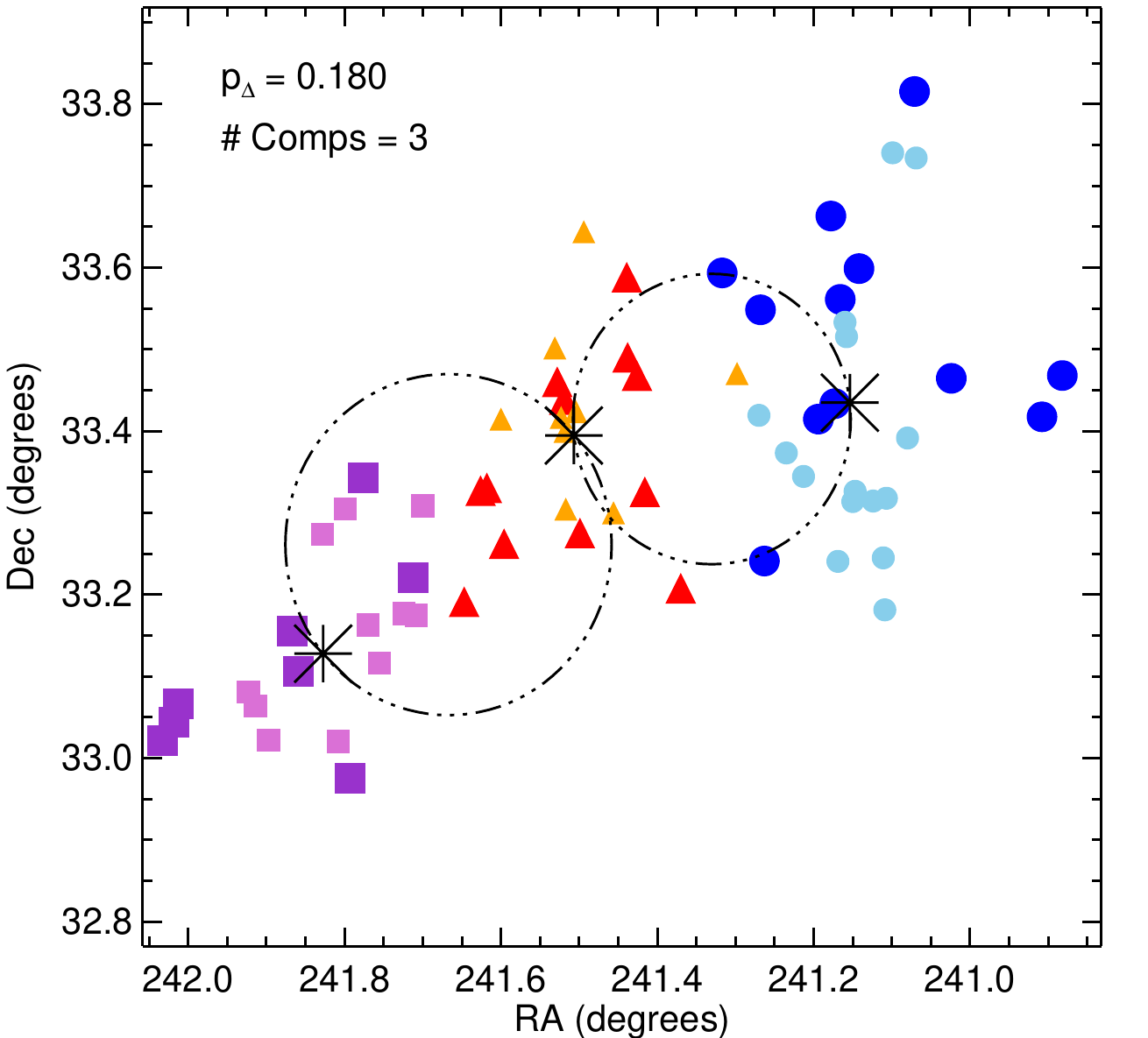}
\includegraphics[scale=0.38]{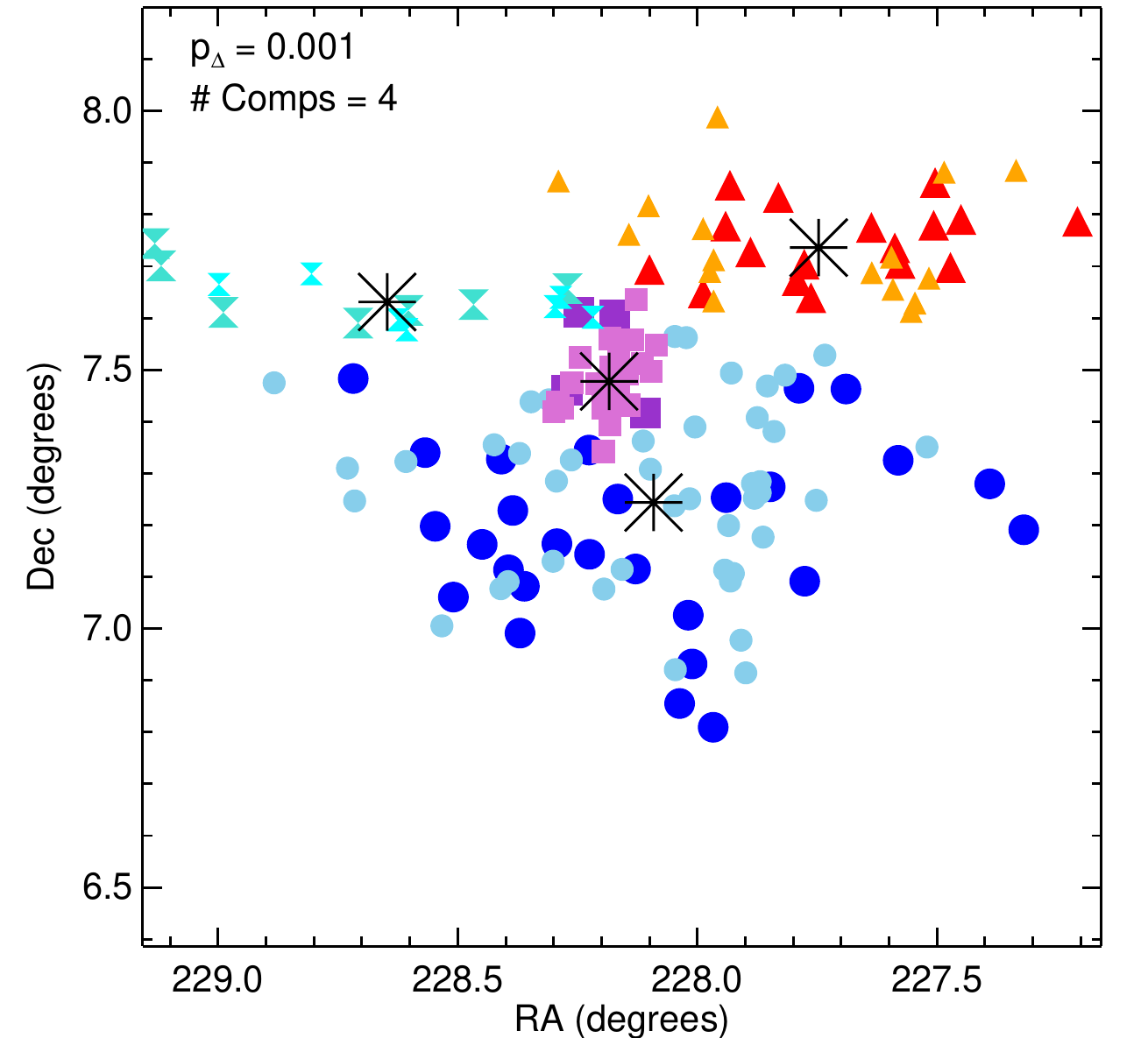}
\includegraphics[scale=0.38]{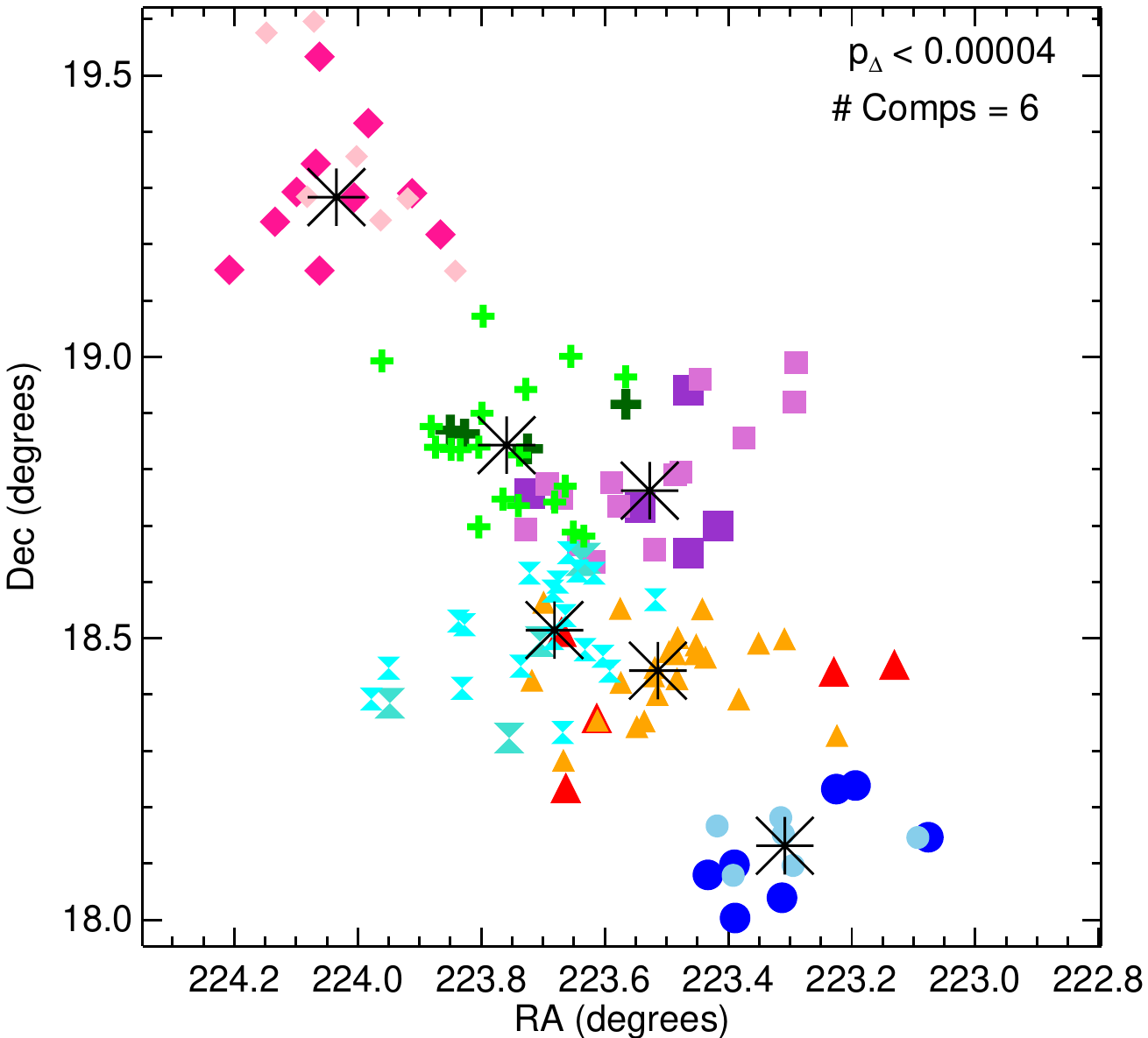}
\includegraphics[scale=0.38]{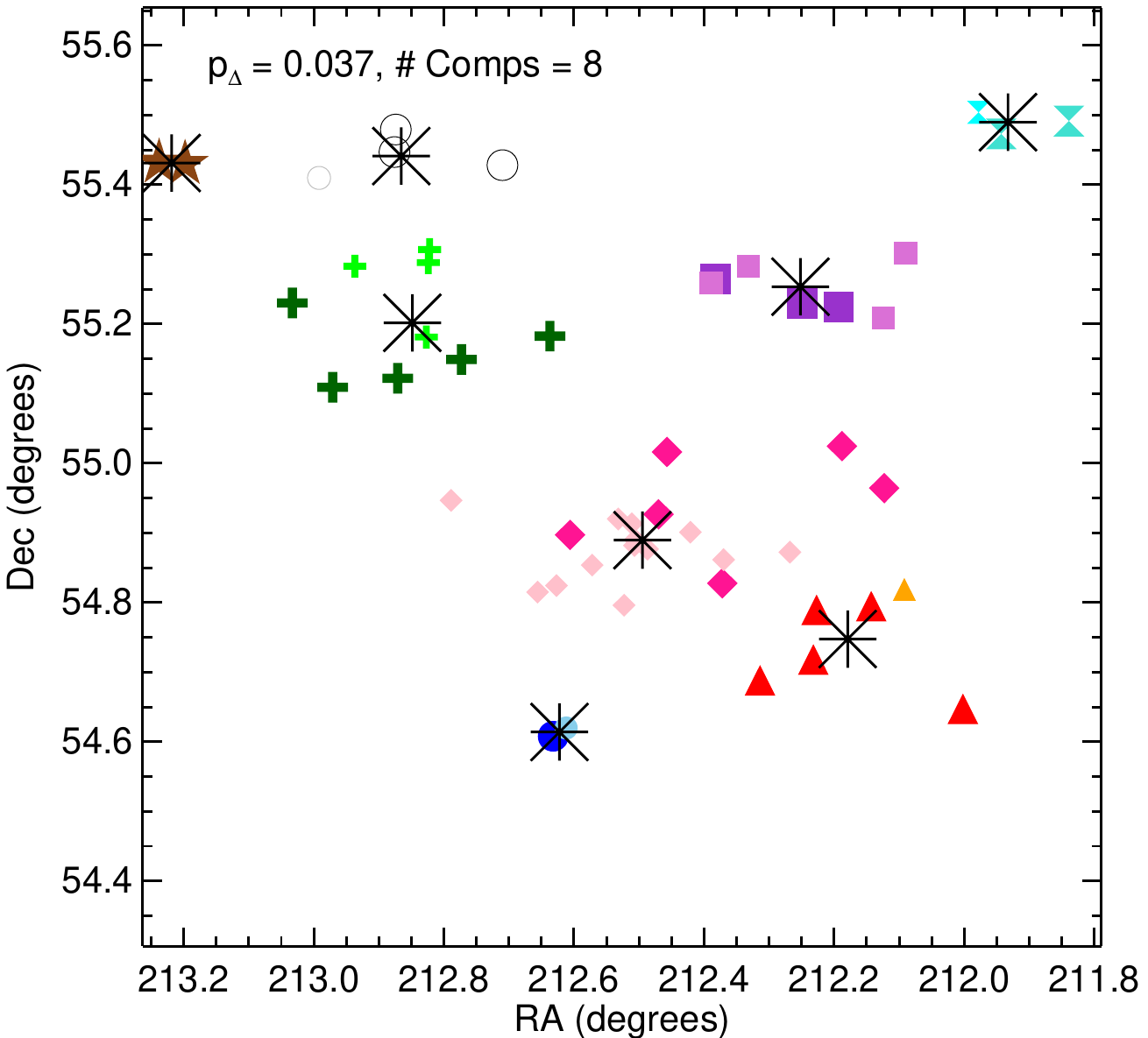}
\caption{Six examples of clusters with different numbers of components: from top left running clockwise, clusters 60539 (A1516), 18, 323, 11015, 62138 (A1991), and 34726 (A2040).  Cluster IDs and matches to Abell clusters are from \citetalias{Einasto2012a}.  North is up and east is to the left.  The small symbols and overall colors (blue, red, purple, turquoise, green, pink, black, and brown) represent galaxies belonging to different components, and dark and light shades of these colors (larger and smaller symbols) correspond to star-forming and non-star-forming galaxies respectively.  The large asterisks indicate centers of components, and the dotted-dashed circles enclose regions between components, as described in \S\ref{sec:BetwMethodSec}.  The corresponding $\Delta$ test $p$-values and number of components are included for each cluster.}
\label{fig:ClustersFig}
\end{center}
\end{figure*}

We use the detection of H$\alpha$ emission as a cut between star-forming and passive galaxies.  We define detection of an emission line if its EW is at least 3 \AA, which is a compromise between the cuts used by \citet[2 \AA]{Rines2005}, \citet[4 \AA]{Balogh2004}, and \citet[5 \AA]{Ma2008}.  A small change in our emission-line detection limit does not significantly change our overall conclusions.  When all four necessary emission lines are detected, we also make use of the BPT diagram \citep{Baldwin1981} that plots $\log([\textnormal{OIII}]\lambda5007/\textnormal{H}\beta)$ vs. $\log([\textnormal{NII}]\lambda6583/\textnormal{H}\alpha)$; the delineating lines from \citet{Kewley2001} and \citet{Kauffmann2003} separate star-forming galaxies from AGN and LINERs \citep[see also, e.g.,][]{Pimbblet2012}, which we exclude from our analysis.

In addition to using H$\alpha$ detections for identifying star-forming galaxies, we also use sSFR as a secondary method to test our analysis.  Compared to emission line EW, sSFR is a tangible, physical property of galaxies that removes any biases related to galaxy mass.  To define our cut between star-forming and passive galaxies, we turn to \citet{Wetzel2012}, who find a clear break in the sSFR distribution of cluster satellite galaxies at $\sim 10^{-11}\, \textnormal{yr}^{-1}$, regardless of stellar mass, total cluster mass, or cluster-centric radius.  We therefore classify star-forming galaxies as those with sSFR $> 10^{-11}\, \textnormal{yr}^{-1}$.  Only about six percent of the classifications of galaxies as star-forming or not differs from our H$\alpha$ to our sSFR method.  Therefore, for simplicity, in our analysis we only use EW classification to detect SF, but separating galaxies by sSFR produces essentially identical conclusions.

As a check on our SF analysis, and to more readily compare our conclusions with past studies \citep[e.g.][]{Tomita1996, Metevier2000, Hou2012}, we also investigate the presence of blue galaxies in our clusters. As described in \citet{Hwang2009}, blue and red galaxies are separated by a line in ${}^{0.1}(u-r)$--$M^{0.1}_{r}$ space defined by \citet{Choi2007} and fit by eye here.  As explained above, the superscript denotes a \emph{K}-correction to a redshift of 0.1.  We find that the blue galaxy fractions in our clusters are equivalent to the star-forming fractions to within a few percent, and using blue fraction as a proxy for SF yields essentially identical results.  Therefore, we only present results using SF fraction in this paper.

Our final sample contains 17 clusters with only one component and 90 clusters with multiple components, determined from tests of substructure performed by \citetalias{Einasto2012a}, as described in \S\ref{sec:SubstructureMethodSec}.  After our absolute magnitude cut, about two thirds of the clusters have stellar masses of about $1\text{--}3 \times 10^{12}\:\textnormal{M}_{\odot}$, and all but two have total stellar masses less than $10^{13}\:\textnormal{M}_{\odot}$.  The two remaining clusters have stellar masses around $10^{13}\:\textnormal{M}_{\odot}$.  Individual galaxies range in stellar mass from about $10^{8}$ to $4 \times 10^{12}\:\textnormal{M}_{\odot}$, though most lie between $10^{10}$ and $10^{11}\:\textnormal{M}_{\odot}$.

Figure~\ref{fig:ClustersFig} illustrates some examples of clusters in the \citetalias{Einasto2012a} catalogue with different numbers of components: from top left running clockwise, clusters 60539 (A1516), 18, 323, 11015, 62138 (A1991), and 34726 (A2040).  Cluster IDs and matches to Abell clusters are from \citetalias{Einasto2012a}.  The different symbols and overall colors represent galaxies belonging to different 3D components as determined by \emph{Mclust}.  Dark and light shades of these colors (larger and smaller symbols) correspond to star-forming and non-star-forming galaxies respectively.  The corresponding $p$-values from the $\Delta$ test, as discussed in \S\ref{sec:TestsMethodSec}, and the number of components in each cluster, are shown in the plots.

These figures give examples of clusters with varying degrees of substructure.  It is interesting to note that the $\Delta$ test $p$-values do not scale directly with the number of cluster components.  For example, the three-component cluster 323 has a higher $p_{\Delta}$, and thus a lower probability of substructure, than either the unimodal cluster 60539 or the bimodal cluster 18.  This illustrates that, when determining substructure, it is important to employ a variety of techniques for comparison.

\section{Methods}
\label{sec:MethodsSec}

\subsection{Measures of Substructure}
\label{sec:SubstructureMethodSec}

We obtain substructure properties of our clusters from \citetalias{Einasto2012a}.  To analyze substructure, they used a number of three-, two-, and one-dimensional tests: multidimensional normal mixture modelling with the \emph{Mclust} package \citep{Fraley2006}; the Dressler-Shectman, also known as the DS or $\Delta$, test \citep{Dressler1988}; the $\alpha$ test \citep{West1990}; the $\beta$ test \citep{West1988}; and a number of one-dimensional tests like the Shapiro-Wilk normality test \citep{Shapiro1965}, the Anderson-Darling test \citep[whose reliability is confirmed by][]{Hou2009}, the Anscombe-Glynn test for kurtosis \citep{Anscombe1983}, and the D'Agostino test for skewness \citep{DAgostino1970}.  The \emph{Mclust} package returns the number of 3D and 2D substructure components in a cluster, and the other tests return $p$-values corresponding to the likelihood that a given cluster contains substructure, with lower $p$-values indicating a higher probability of substructure.

In this study, we use four of these measures of substructure: mixture modelling via the \emph{Mclust} package, the $\Delta$ test, the $\alpha$ test, and the $\beta$ test.  We briefly explain each here.  More details on the tests can be found in \citetalias{Einasto2012a} and references therein.

\subsubsection{The \emph{Mclust} Package}
\label{sec:MclustMethodSec}

To search for possible components in clusters, \citetalias{Einasto2012a} applied multidimensional normal mixture modelling based on the analysis of a finite mixture of distributions, in which each mixture component is taken to correspond to a different group, cluster, or subpopulation.  To model the collection of components, they used the \emph{Mclust} package for classification and clustering \citep{Fraley2006} from \emph{R}, an open-source free statistical environment developed under the GNU GPL \citep[][\texttt{http://www.r-project.org}]{Ihaka1996}.  This package searches for an optimal model for the clustering of the data among models with varying shape, orientation, and volume, and finds the optimal number of components and the corresponding classification (i.e., the membership of each component).

For every galaxy, \emph{Mclust} calculates the probabilities of it belonging to any of the components.  The uncertainty of classification for each galaxy is defined as one minus the highest probability of that galaxy belonging to a component.  The mean uncertainty for the full sample is used as a statistical estimate of the reliability of the results.  Thus, using this information, we are able to assign component membership to each cluster galaxy and determine the number of components in each cluster.

\subsubsection{The $\Delta$, $\alpha$, and $\beta$ Tests}
\label{sec:TestsMethodSec}

The Dressler-Schectman (DS or $\Delta$) test \citep{Dressler1988} calculates how discrepant each galaxy's local velocity mean and dispersion are from the cluster values.  Each galaxy is assigned a value $\delta_{i}$ measuring this deviation, with higher values corresponding to stronger local deviations.  The sum of these values over all cluster members, $\Delta = \Sigma\delta_{i}$, is a quantitative measure of the cluster substructure.

To determine the significance of this value, these results were calibrated by \citetalias{Einasto2012a} using Monte Carlo simulations, in which the velocities of the galaxies were randomly shuffled among the positions.  This procedure effectively destroys any true correlation between velocities and positions.  For each cluster, \citetalias{Einasto2012a} ran 25000 simulations and each time calculated $\Delta_\mathrm{sim}$.  The significance of having substructure (the $p$-value) can be quantified by the ratio $p = N(\Delta_\mathrm{sim} > \Delta_{obs})/N_\mathrm{sim}$ -- the ratio of the number of simulations in which the value of $\Delta_\mathrm{sim}$ is larger than the observed value, to the total number of simulations.  A smaller $p$-value corresponds to fewer simulated clusters having higher $\Delta_\mathrm{sim}$ values than the true cluster.  Thus, smaller $p$-values indicate higher probabilities of substructure.

The $\alpha$ test \citep{West1990} measures the centroid shift for each galaxy's local neighborhood, searching for regions that show correlations between galaxy positions and velocities that differ from those of the whole cluster.  Each galaxy is assigned a value $\alpha_{i}$ as a measure of its centroid shift.  The average of all $\alpha_{i}$ values is defined as the cluster's $\alpha$-value, a measure of how much the centroid of all galaxies shifts as a result of local correlations between positions and velocities.

The 2D $\beta$ test, presented in \citet{West1988}, is a test for the asymmetry in the sky distribution of galaxies in groups.  Each galaxy's asymmetry, $\beta$, is calculated, and the average asymmetry of the whole population, $\langle \beta \rangle$, measures the possible substructure in the cluster.  According to \citet{Pinkney1996}, this test is sensitive to mirror asymmetry, but not to deviations in radial symmetry.

\begin{figure*}
\begin{center}
\includegraphics[scale=0.58]{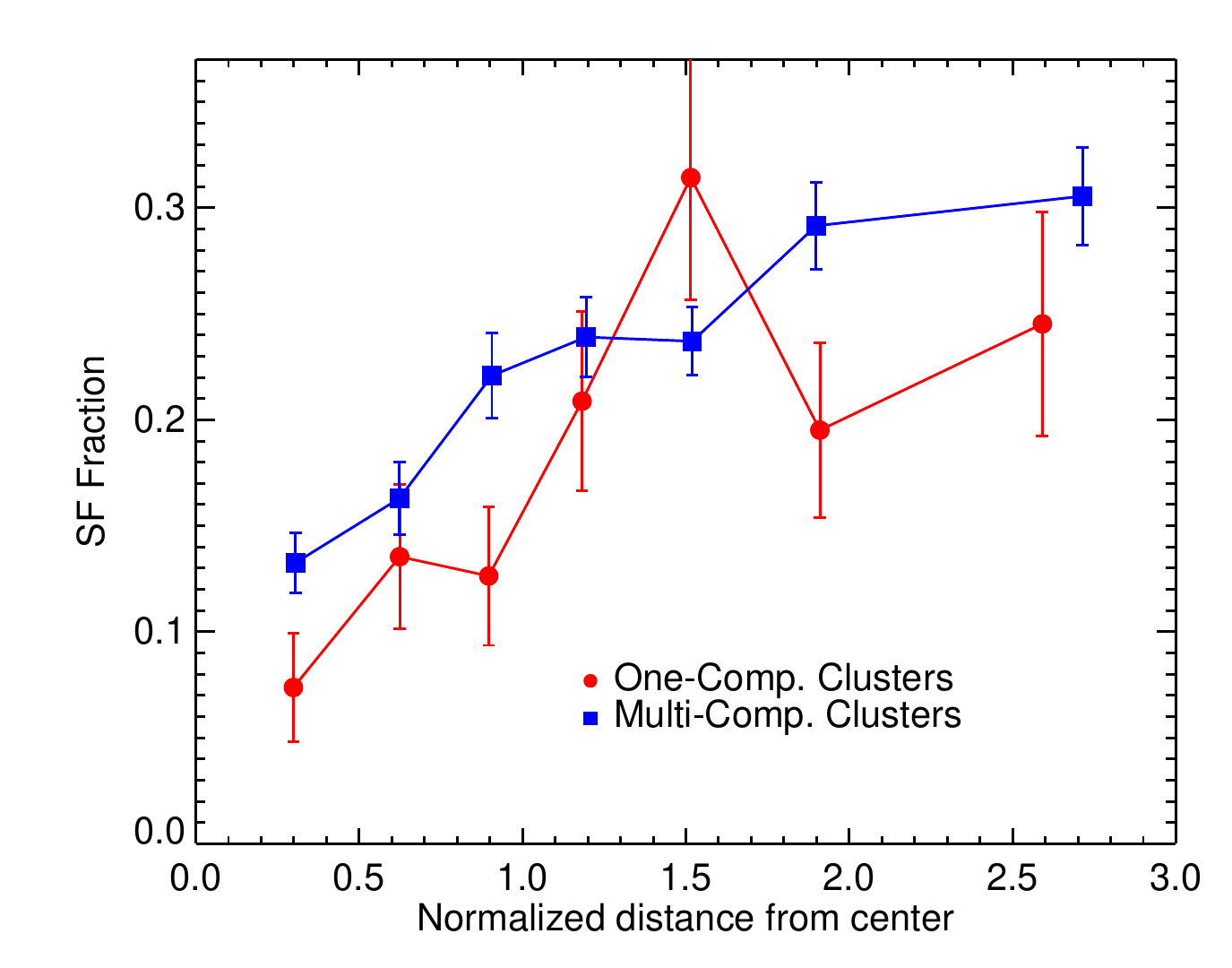}
\includegraphics[scale=0.58]{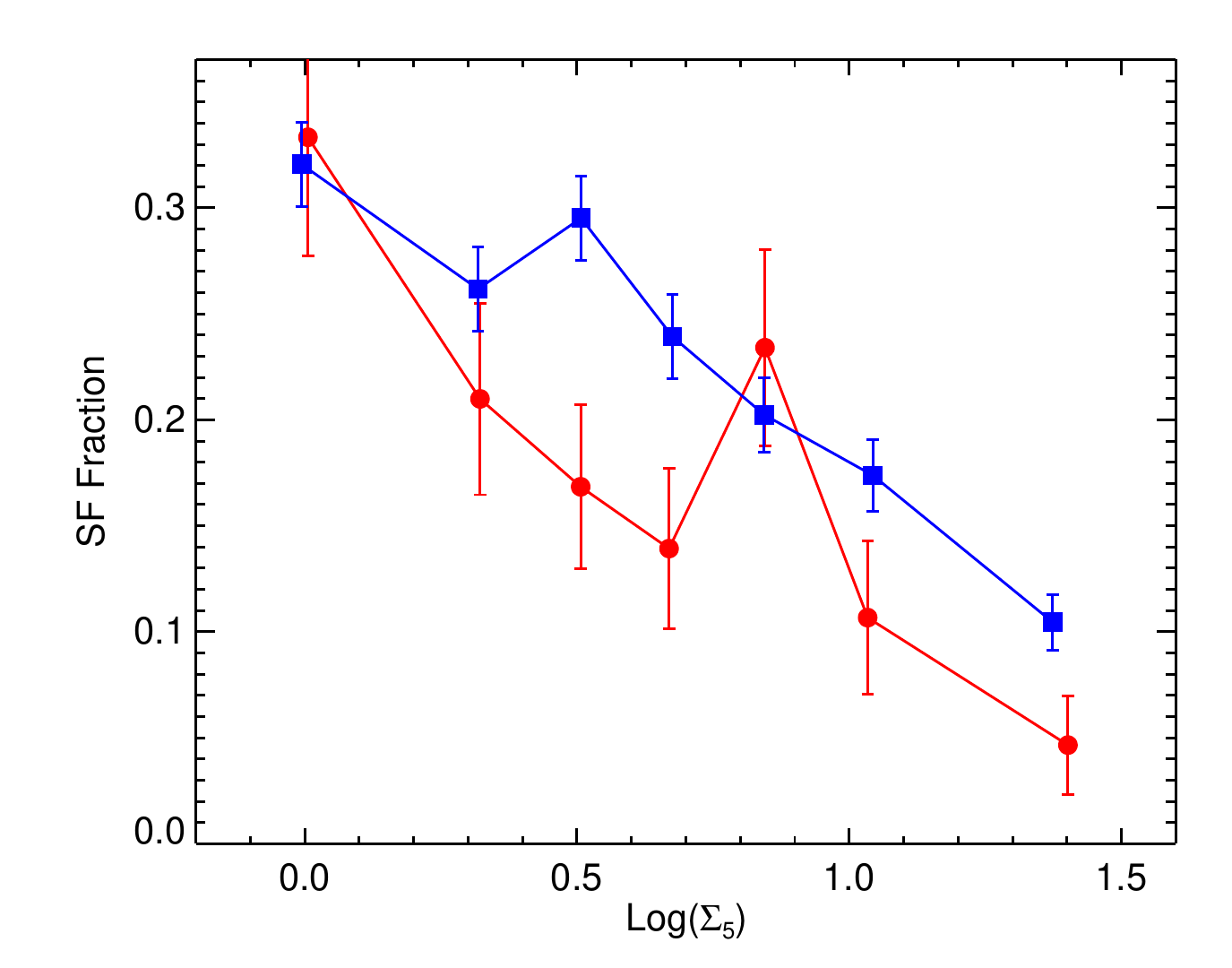}
\caption{SF fraction vs. normalized radial distance (left) and local number density (right), in single-component clusters (red circles) and multi-component clusters (blue squares).  Points represent binned SF averages.  The higher SF in multi-component clusters than single-component clusters at most distances and densities indicates that merging is related to more SF.}
\label{fig:SFvsDistDensFig}
\end{center}
\end{figure*}

The $\alpha$ and $\beta$ tests were both calibrated by \citetalias{Einasto2012a} in the same manner as the $\Delta$ test.  In these test as well, smaller $p$-values indicate higher probabilities of substructure.  Thus, the $p$-values from these tests provide quantitative measures of the probability of substructure in our clusters.

\subsection{Local Density and SF Determination}
\label{sec:DensMethodSec}

As a proxy for local number density of galaxies, we assign each galaxy a value $\Sigma_{n}=n/\pi d_{n}^{2}$, a widely-used local density estimate based on $d_{n}$, the projected physical distance to the galaxy's $n^{\textnormal{th}}$-nearest neighbor \citep[e.g.,][]{Balogh2004, Rines2005}.  Following a common methodology, we use $n = 5$ for all clusters.  The local SF fraction around each galaxy is defined out to the same $5^{\textnormal{th}}$-nearest neighbor as is used to calculate local density.  Using other values of $n$, such as $n = 10$ or $n = \sqrt{N}$ where $N$ is the number of member galaxies in a cluster, does not significantly affect our results.

\subsection{Regions Between Centers of Components}
\label{sec:BetwMethodSec}

In \S\ref{sec:SFvsCompsSec} we investigate the presence of SF in regions between the centers of interacting components.  Therefore, we must determine what it means for a galaxy to be ``between" component centers.  We choose to investigate these regions in particular because past studies have argued for SF enhancement, or lack thereof, specifically in the regions between the centers of merging subclusters.  However, these studies used different demarcations of these regions.  \citet{Hwang2009}, studying the clusters A168 and A1750, defined one subcluster in each cluster as a ``hub" around which a line could be swept.  The line connecting the subclusters was defined as zero degrees.  The number of emission-line galaxies was then counted as a function of angle from this connecting line.  \citet{Cortese2004}, \citet{Ferrari2005}, and \citet{Wegner2011} argued for enhanced SF between components using visual inspection, confirmed by statistical tests of the difference between the distributions of emission-line and non-emission-line galaxies  (such as the 2D Kolmogorov-Smirnovov test).

We choose to use a different definition of the region between component centers in an attempt to be as consistent and robust as possible across our entire sample.  We first use the biweight location estimator of \citet{Beers1990} to determine the centers of each component.  We then define the regions between component centers as circles whose diameters span the distance between the component centers, as illustrated by the dotted-dashed circles in Figure~\ref{fig:ClustersFig}.  This is a quantitative approximation of the various regions between components used in the literature.  Galaxies found inside these circles are defined as being between component centers.  In this way, our regions between component centers include galaxies near to the components' connecting line, but ignores galaxies that lie unreasonably far away from this line.

It is important to note that every galaxy in our sample belongs to a component, including galaxies defined as being between component centers.  These galaxies between component centers are defined as such because they reside in regions that, in individual clusters studied in the past, have been shown specifically to contain enhanced SF.  To be thorough, in addition to our definition above, we also tested different definitions of our regions between component centers: circles whose diameters are three-quarters or half the distance between component centers; ellipses whose major axes span the distance between component centers and whose minor axes are three-quarters or half this distance; and the same ellipses turned ninety degrees.  In all cases, we find that our conclusions in \S\ref{sec:SFvsCompsSec} remain unchanged.

To simplify the interpretation of our results, in this section we include only on those clusters with two or three components.  For clusters with three components, we mostly focus only on the two outer component pairs to avoid the complication of three overlapping regions.  An exception to this is when the three component centers describe a rough equilateral triangle.  In this case, we include all three component pairs in our analysis, as this arrangement of components does not produce significantly overlapping regions.  This slight discrepancy does not affect our conclusions, as there are only seven tri-modal clusters out of 28 for which we included all three component pairs.

\section{Results}
\label{sec:ResultsSec}

To determine the relationship between cluster substructure/merger history and SF, we compare several measures of SF with various measures of substructure and environment.  In \S\ref{sec:SFvsDistDensSec} we measure the relationship between SF and clustercentric distance, and SF and number density, and compare the results between single- and multi-component clusters; in \S\ref{sec:SFvsCompsSec} we investigate SF and number density in regions between components; and in \S\ref{sec:SFvsClusterSec} we analyze SF fraction against statistical measures of substructure.

\subsection{Star Formation in Single- and Multi-Component Clusters}
\label{sec:SFvsDistDensSec}

In this section, we compare SF fractions in single- and multi-component clusters by investigating the well-established direct correlation between SF and radial distance, and the inverse correlation between SF and number density \citep[e.g.,][]{Dressler1980, Depropris2004, Kodama2004, Rines2005}.

Our results are shown in Figure~\ref{fig:SFvsDistDensFig}, which includes information on single-component clusters only (red circles) and multi-component clusters only (blue squares).  We plot SF fraction as a function of normalized radial distance (on the left) and local number density (on the right).  Each point represents the total fraction of galaxies that are star-forming in a given distance or density bin across all clusters, and the $1\sigma$ uncertainties are calculated via a bootstrap resampling of the SF in the galaxies in each bin.  We normalize each galaxy's radial distance by the virial radius of its host cluster.  Each cluster's center is determined using the biweight location estimator of \citet{Beers1990}, and virial radii are supplied by \citetalias{Einasto2012a}.  Local density determination is described in \S\ref{sec:DensMethodSec}.

In general, we reproduce the expected direct correlation between SF and radial distance, and the inverse correlation between SF and density.  This result is expected for single-component clusters, as no effects from merging could have disrupted the observed distribution.  Interestingly, in multi-component clusters, we find that these correlations are still preserved.  Indeed, \citet{Ribeiro2013} find a decreasing fraction of red galaxies with distance from the centers of both Gaussian and non-Gaussian clusters.

The fraction of star-forming galaxies in all multi-component clusters is $0.228 \pm 0.007$, higher than the SF fraction in single-component clusters, $0.175 \pm 0.016$, a difference of $3\sigma$.  These $1\sigma$ uncertainties are calculated via a bootstrap resampling of the SF galaxies in the clusters.  Furthermore, multi-component clusters exhibit higher SF fractions at almost all distances and densities than single-component clusters.  The spike in SF fraction in single-component clusters around a normalized distance of 1.5 is caused by two highly asymmetrical, elongated clusters with unusually high SF fractions at that distance.  When classifying star-forming galaxies using sSFR as described in \S\ref{sec:DataSec}, we obtain a very similar and equally significant difference between the average SF fractions in single- and multi-component clusters.  These results suggest that, in general, more SF is found in merging clusters than relaxed clusters.

It is unclear whether cluster mass affects the amount of observed SF.  For example, several studies \citep[e.g.,][]{Finn2005, Homeier2005, Poggianti2006, Koyama2010} find an inverse correlation between SF and cluster mass, while others \citep[e.g.,][]{Goto2004, Popesso2007, Balogh2010, Chung2011} assert that no such correlation exists.  We therefore check whether the difference in SF fraction between single- and multi-component clusters could be due to a difference in cluster mass.  For these purposes total stellar mass ($\textnormal{M}_{*}$) is a reasonable proxy for total halo mass, given the tight observed relation between these quantities \citep[e.g.,][]{Andreon2010, Gonzalez2013}.  To accomplish this, we weight the galaxies to have the same $\textnormal{M}_{*}$ distribution for both the single- and multi-component clusters, performing the following steps for each distance and density bin.  We first determine the distribution, normalized to unity, of the stellar masses of the galaxies in all clusters.  We then determine this distribution for single- and multi-component clusters separately.  Next, we weight the stellar masses of the galaxies in single- or multi-component clusters so their normalized distribution matches that of all galaxies in the bin.  Finally, we define the weighted SF fraction as the total of the weights of star-forming galaxies divided by the total of the weights of all galaxies.  The effect of this normalization is to remove any effect of stellar mass on the SF fraction measurements.

We find that the weighted SF fractions in each distance and density bin only differ by a few percent from the unweighted SF fractions, preserving the observed trends in distance and density.  This is also true of the weighted SF fractions in single- and multi-component clusters in general.  If anything, the normalization very slightly increases the significance of this difference.  We therefore conclude that stellar mass is not the cause of the observed difference in SF between single- and multi-component clusters.

We also investigate the difference in average sSFR between single- and multi-component clusters.  We compute both the average of the sSFRs in each set of galaxies, defined as the linear average, and the average of the logarithms of the sSFRs in each set of galaxies, defined as the log-average.  The log-average is calculated because the sSFRs span a wide range of values in log space.  Between single- and multi-component clusters, the differences in the linear averages and log-averages exhibit only $1.2\sigma$ and $1.7\sigma$ significance, respectively, though qualitatively they behave similarly to the difference in SF fraction above.

The remainder of the paper will use SF fraction as calculated using the detection of H$\alpha$ as an SF indicator, as described in \S\ref{sec:DataSec}, for the following reasons.  First, the calculation of SF fraction relies on only a single observable -- equivalent width of the H$\alpha$ emission line -- while determining sSFR introduces errors inherent in estimating stellar mass.  Second, the average sSFR in both single- and multi-component clusters is more than $3\sigma$ below our demarcation between star-forming and passive galaxies ($\textnormal{sSFR} = 10^{-11}\, \textnormal{yr}^{-1}$).  Because of this, the calculation of average sSFR is dominated by passive galaxies, and changing the number of star-forming galaxies has little effect on this value.  We therefore choose to use SF fraction as the simplest and most robust measure of SF.  We note that it is promising that the behavior of average sSFR in single- and multi-component clusters is qualitatively similar to the behavior in SF fraction, though the significance of this difference is lower.

\subsection{Star Formation and Density Between Components}
\label{sec:SFvsCompsSec}

To more definitively study where the higher fraction of SF might reside in multi-component clusters, in this section we investigate the presence of star-forming galaxies in the regions between centers of cluster components, since a number of studies show an enhancement of SF specifically in the regions between two merging subclusters \citep[e.g.,][]{Cortese2004, Ferrari2005, Hwang2009, Wegner2011}.  We seek to determine whether the higher SF we observe in our multi-component clusters in the previous section might be especially concentrated in regions between component centers.  We compare the SF and density properties of regions between component centers to regions elsewhere.  As discussed in \S\ref{sec:BetwMethodSec}, we only include clusters with two or three components in this analysis, and changing the definition of the regions between component centers does not affect our conclusions.

\begin{figure}
\begin{center}
\includegraphics[scale=0.67]{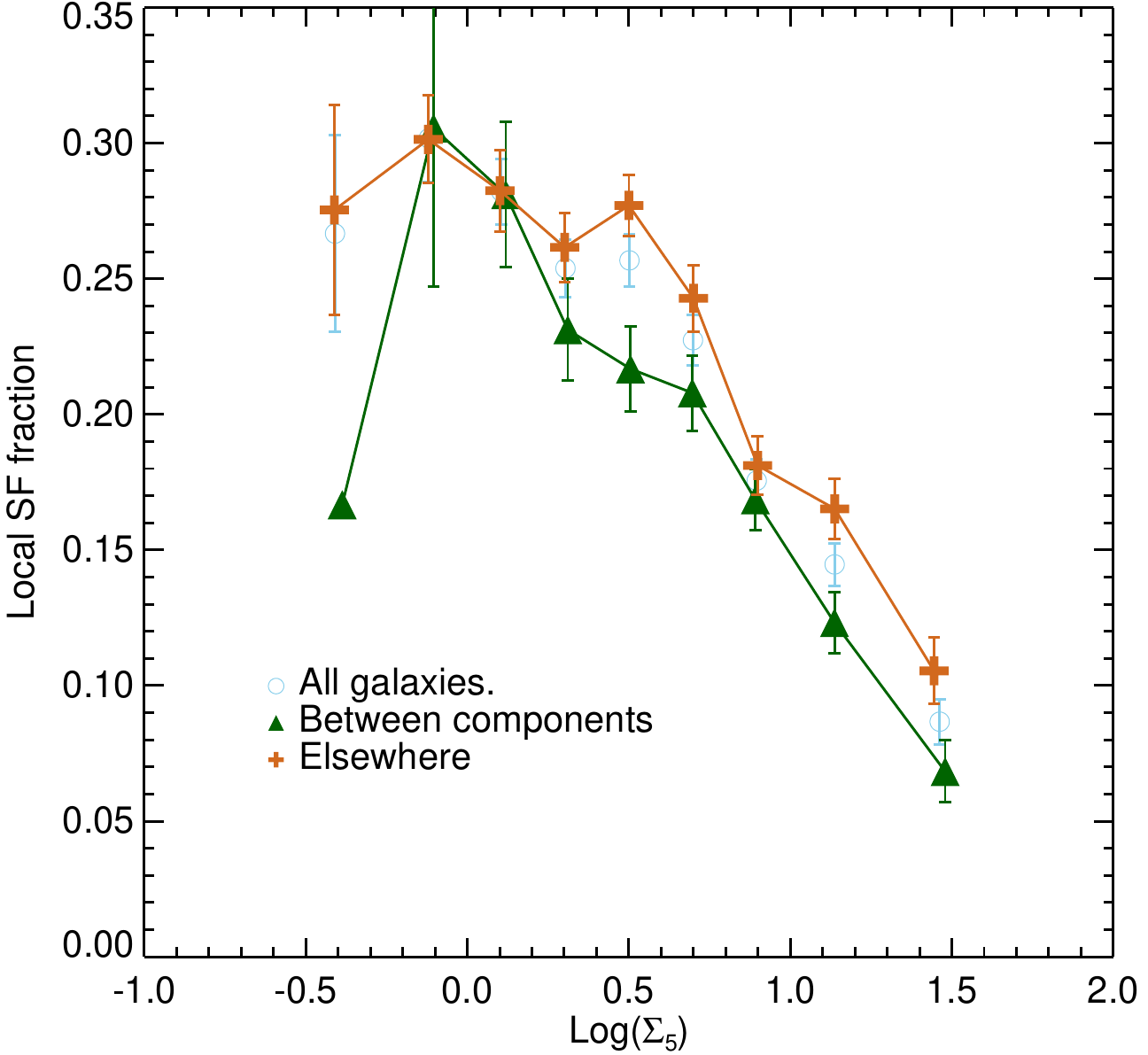}
\caption{Local SF fraction vs. $\Sigma_{5}$ for all galaxies (open light blue circles), galaxies between components (dark green triangles), and galaxies elsewhere (light brown plus signs).  At almost all densities, we find slightly less SF in regions between components than in regions elsewhere.}
\label{fig:SFfracvsDensFig}
\end{center}
\end{figure}

Figure~\ref{fig:SFfracvsDensFig} plots local SF fraction as a function of local density for all galaxies (open light blue circles), galaxies between component centers (dark green triangles), and galaxies elsewhere (light brown plus signs).  Each point represents the average local SF fraction found in that density bin, and the $1\sigma$ uncertainties are calculated via a bootstrap resampling of the local SF fractions of the galaxies in each bin.

At most densities, we find slightly lower average SF fractions in regions between component centers than in regions elsewhere.  At those densities where this is not the case, the two regions exhibit equivalent average SF fractions.  The very low SF fraction observed between component centers at the lowest density is due to the fact that very few regions between component centers in our sample exhibit densities this low.  In general, the average local SF fraction in regions between component centers is $0.177 \pm 0.006$, lower than the average fraction elsewhere, $0.235 \pm 0.005$.  This difference is consistent with the higher galaxy number densities exhibited in regions between component centers compared to regions elsewhere, since more SF is found in regions of clusters with higher density.  These results show that, on average, SF is not enhanced in regions between component centers.

\subsection{Star Formation vs. Cluster Substructure}
\label{sec:SFvsClusterSec}

We wish to determine whether the higher SF fraction in multi-component clusters described in \S\ref{sec:SFvsDistDensSec} increases as the amount of substructure increases.  We therefore investigate the relationship between SF and the amount of cluster substructure, employing the $p$-values from the $\Delta$ test, $\alpha$, and $\beta$ tests, and the number of components in three dimensions as determined by \emph{Mclust}.  We examine the relationship between these $p$-values and the total fraction of star-forming galaxies in the cluster.

\begin{figure}
\begin{center}
\includegraphics[scale=0.67]{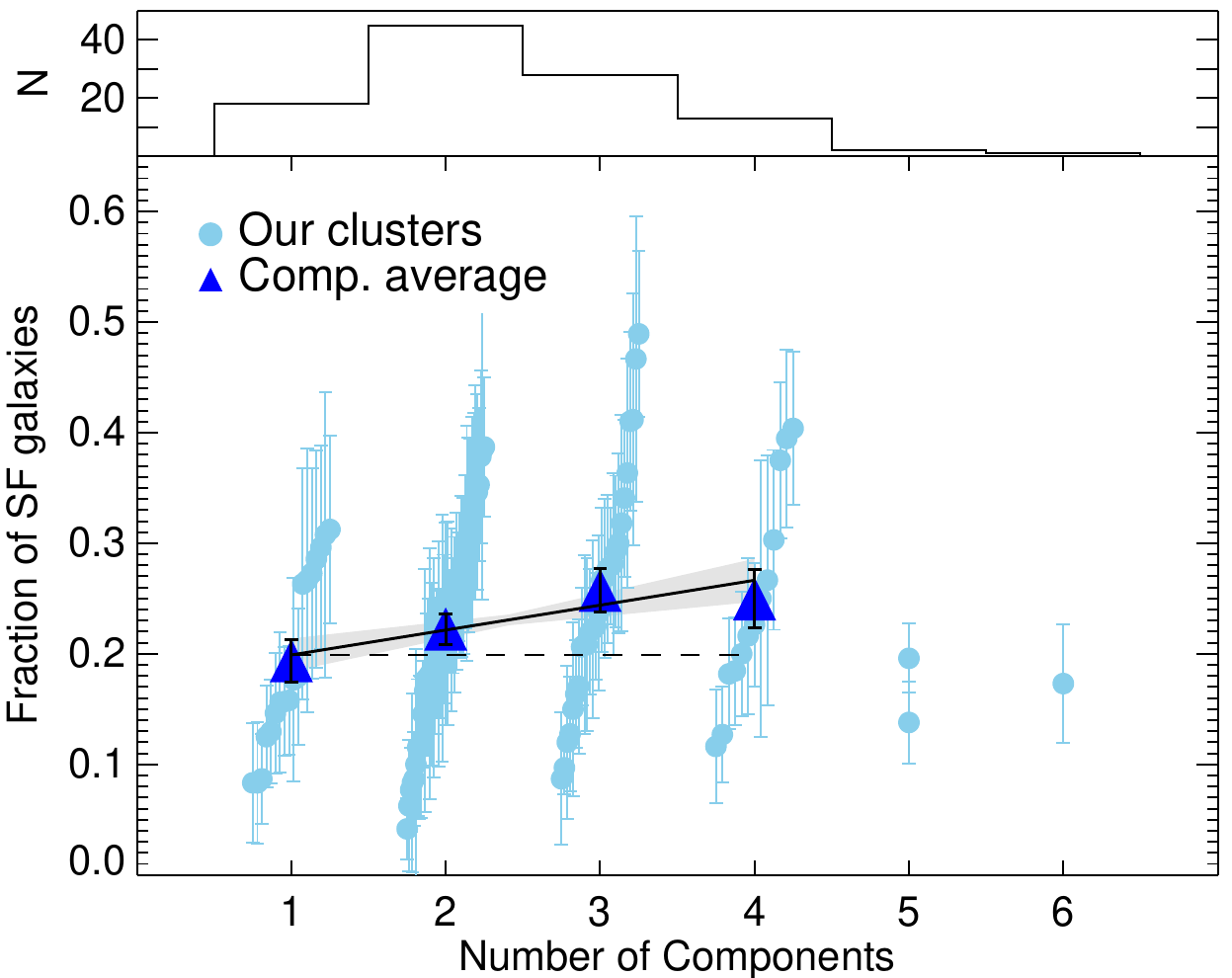}
\caption{SF fraction vs. number of substructures as defined by the \emph{Mclust} procedure from \citetalias{Einasto2012a}.  Light blue circles are our clusters, and dark blue triangles are averages, for each number of components.  Points corresponding to clusters with the same number of components have been slightly offset to the left and right of their true positions to more clearly show error bars.  The gray region represents a $1\sigma$ error on the best fit solid line, for which we ignore the five- and six-component clusters, as explained in the text.  For comparison, the horizontal dashed line illustrates no correlation.  The histogram above the plot shows the number of clusters with different numbers of components.  A weak direct correlation between SF fraction and number of components is found.}
\label{fig:SFvsNumCompsFig}
\end{center}
\end{figure}

\begin{figure*}
\begin{center}
\includegraphics[scale=0.45]{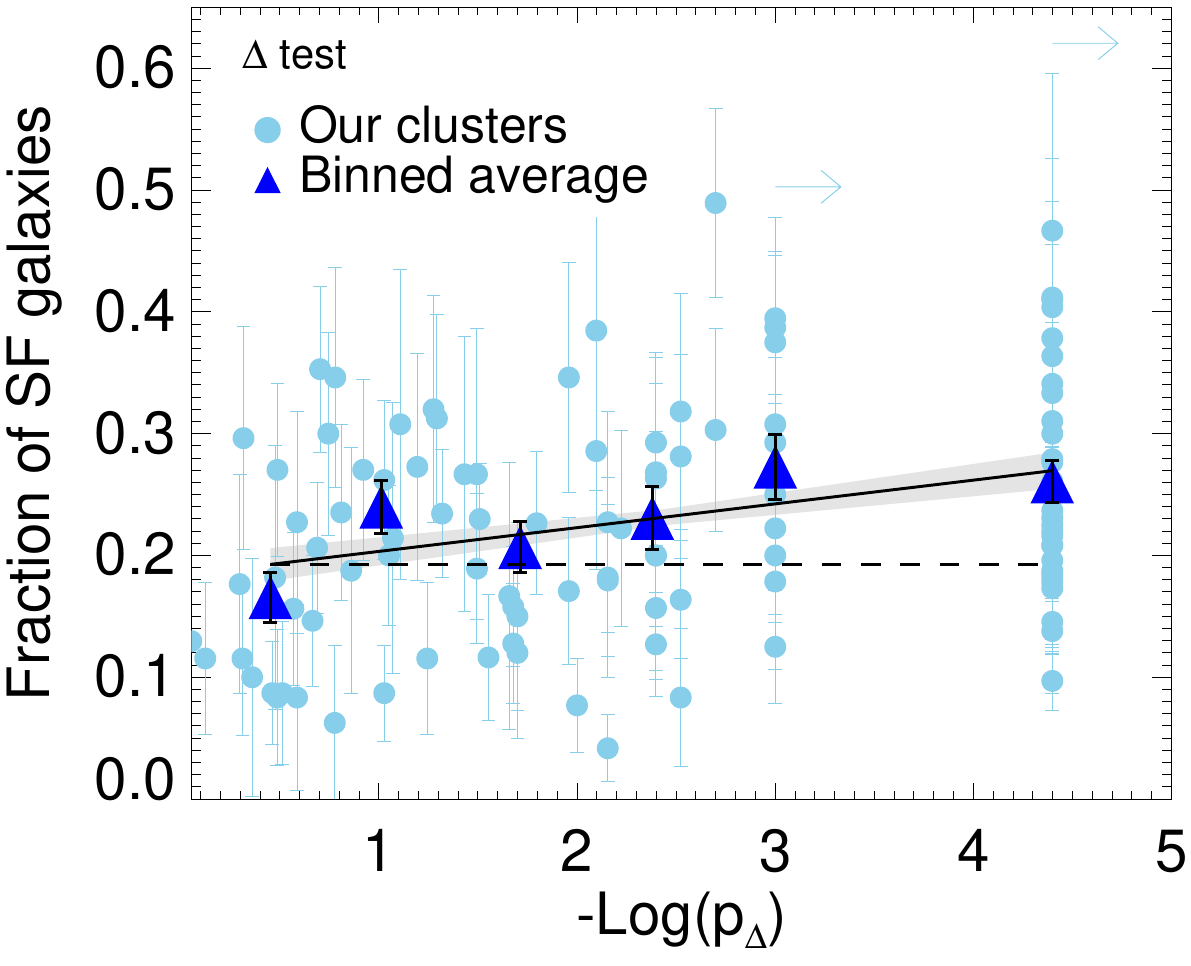}
\includegraphics[scale=0.45]{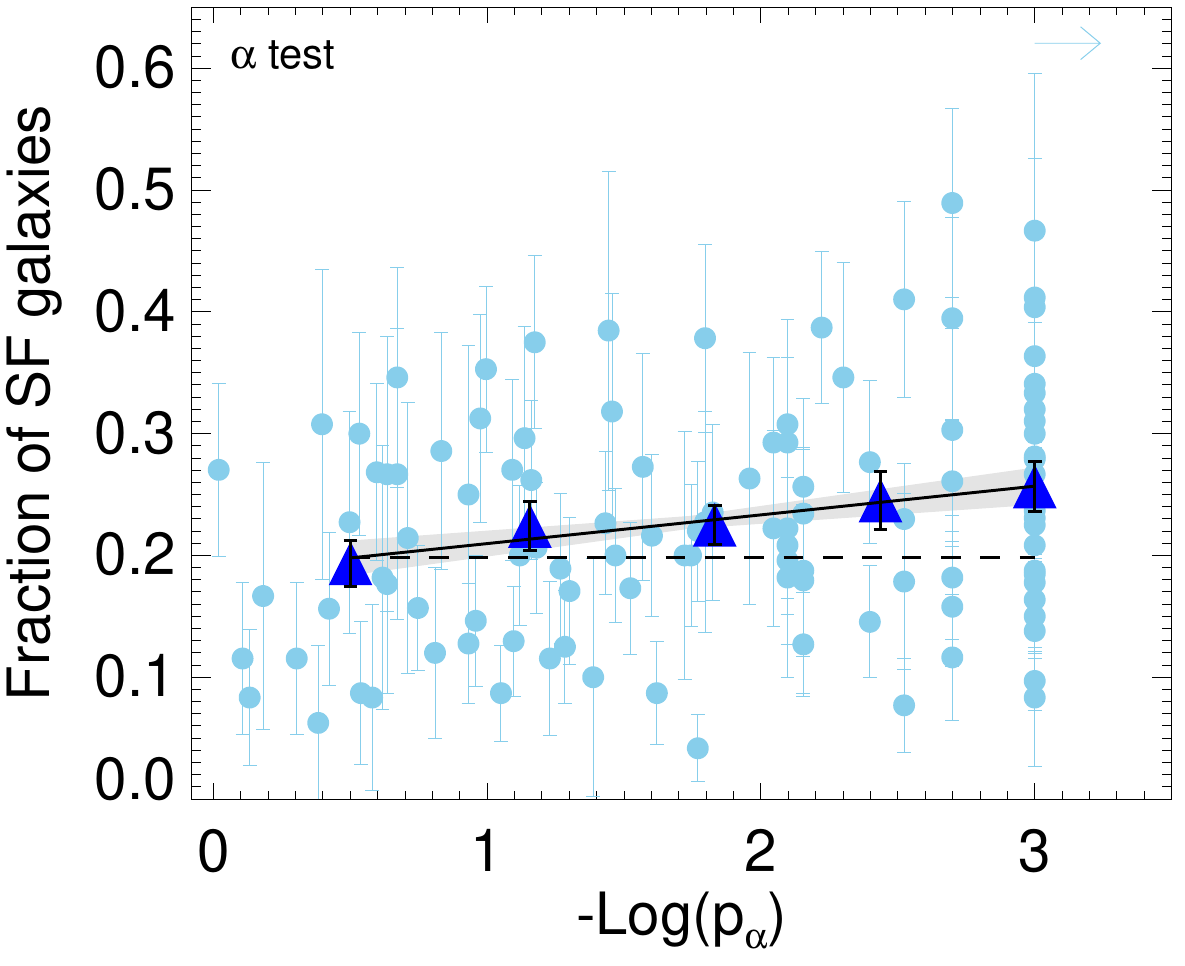}
\includegraphics[scale=0.45]{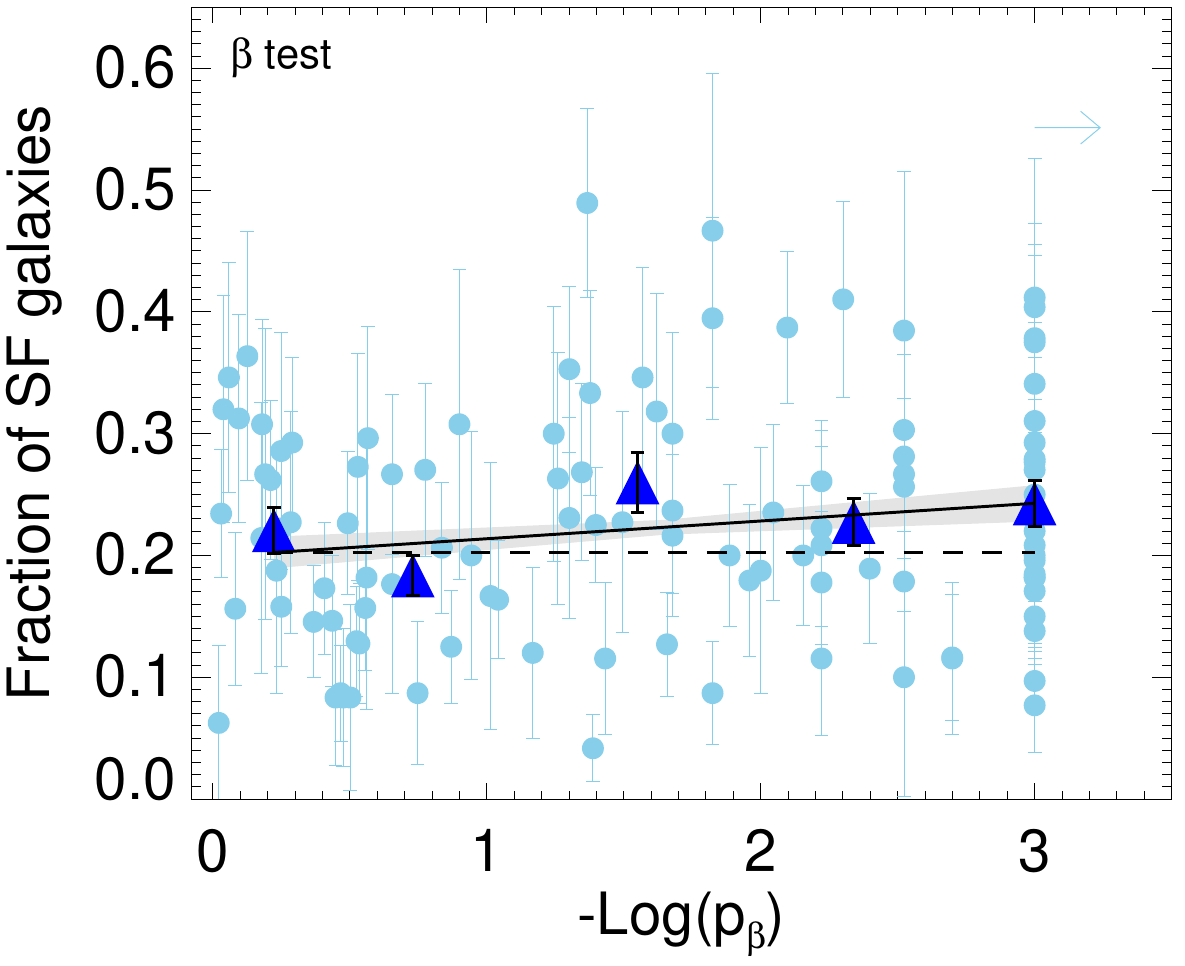}
\caption{SF fraction vs. \emph{negative} $p$-values from (left to right) the $\Delta$ test, the $\alpha$ test, and the $\beta$ test from \citetalias{Einasto2012a}.  Light blue circles are our clusters, and dark blue triangles are binned averages.  The gray regions represent $1\sigma$ errors on the best fit solid lines.  For comparison, the horizontal dashed lines illustrate no correlation.  A weak direct correlation between SF fraction and amount of substructure is found.}
\label{fig:SFfracvspFig}
\end{center}
\end{figure*}

Figures~\ref{fig:SFvsNumCompsFig} and~\ref{fig:SFfracvspFig} plot SF fraction in each cluster as a function of global substructure.  Figure~\ref{fig:SFvsNumCompsFig} shows the fraction of star-forming galaxies as a function of the number of components in each cluster as determined by \emph{Mclust} from \citetalias{Einasto2012a}.  Light blue circles represent individual clusters, and the uncertainties are $1\sigma$ standard deviations calculated via a bootstrap resampling of the SF fractions in each cluster.  Dark blue triangles represent bin averages for which the uncertainties are $1\sigma$ standard errors of each bin.  Points corresponding to clusters with the same number of components have been slightly offset to the left and right of their true positions to more clearly show error bars.  The histogram above the plot shows the number of clusters with different numbers of components.

Due to our absolute magnitude cut, a small number of clusters now contain fewer subclusters than before the cut, since some subclusters only contained galaxies fainter than our absolute magnitude limit.  In this figure, we report the number of components after the cut.  Only three clusters contain more than four components, and the large scatter in the other bins suggests that these three points may not accurately represent the average SF fractions in clusters with more than four components.  Therefore, we do not include these clusters in our fitting, though the effect of including them is discussed below.

Figure~\ref{fig:SFfracvspFig} replaces the number of cluster components with the $p$-values of, from left to right, the $\Delta$ test, the $\alpha$ test, and the $\beta$ test.  Notice that we plot the \emph{negative} logarithm of the $p$-values so that the probability of substructure increases to the right.  Dark-colored triangles are binned averages.  The vertical lines of data points at $p = 0.001$ and $0.00004$ represent clusters whose $p$-values (graph placements) are upper (lower) limits.  In other words, each of these clusters has a very high probability of substructure.  When performing fits to the data, explained below, we assume that these points are exactly where they are displayed.  We discuss below the effects of removing these points from the fits.

\begin{deluxetable*}{lccccc}
\tablecolumns{6}
\tablewidth{0pc}
\tablehead{
\colhead{} & \multicolumn{2}{c}{\textbf{Slopes}} & \colhead{} & \colhead{} & \colhead{}\\
\cline{2-3} \\
\colhead{\phm{spacespac}\textbf{Plot}}\phm{spacespa} & \colhead{\textbf{Unweighted}} & \colhead{\textbf{Weighted by Mass}} & \colhead{$\chi^{2}_{\nu}$\tablenotemark{a}} & \colhead{$\rho$\tablenotemark{b}} & \colhead{\textbf{P}\tablenotemark{c}} \\
\colhead{$(1)$} & {$(2)$} & {$(3)$} & {$(4)$} & {$(5)$} & {$(6)$}}
\startdata
SF frac. vs. \# Comps. & $0.022 \pm 0.010$ & $0.022 \pm 0.011$ & 0.19 & 0.15 & 0.116 \\
SF frac. vs. $p_{\Delta}$ & $0.020 \pm 0.006$ & $0.025 \pm 0.006$ & 1.59 & 0.32 & 0.001 \\
SF frac. vs. $p_{\alpha}$ & $0.024 \pm 0.010$ & $0.026 \pm 0.010$ & 0.10 & 0.24 & 0.013 \\
SF frac. vs. $p_{\beta}$ & $0.015 \pm 0.008$ & $0.022 \pm 0.008$ & 1.48 & 0.09 & 0.359
\enddata
\tablenotetext{a}{Reduced chi-squared statistic for unweighted data.}
\tablenotetext{b}{Spearman's rank coefficient for unweighted data.}
\tablenotetext{c}{Statistical significance of $\rho$.}
\label{tab:FitsTab}
\end{deluxetable*}

As in \S\ref{sec:SFvsClusterSec}, we test any possible effect of cluster mass on SF.  In Figures~\ref{fig:SFvsNumCompsFig} and~\ref{fig:SFfracvspFig} we also calculate the binned averages weighted by the total stellar masses of our clusters.  It is important to note that we do not simply weight each cluster by its total stellar mass; rather, our normalization procedure, similar to that in \S\ref{sec:SFvsClusterSec}, is as follows.  We first calculate each cluster's stellar mass, defined as the total stellar mass of all member galaxies.  We then determine the distribution, normalized to unity, of these cluster masses in bins of $10^{12}\:\textnormal{M}_{\odot}$.  Next, for each bin in Figures~\ref{fig:SFvsNumCompsFig} and~\ref{fig:SFfracvspFig}, we weight the stellar masses of the clusters in the bin so their normalized distribution matches that of our entire sample.  Finally, we apply these weights to the SF fraction measurements of the clusters.  The effect of this normalization is to prevent a few unusually high- or low-mass clusters from artificially affecting the average in any particular bin.  For clarity, we have only shown the unweighted averages in our plots.

Figures~\ref{fig:SFvsNumCompsFig} and~\ref{fig:SFfracvspFig} include solid lines of best fit to the binned averages, with their $1\sigma$ errors represented by the shaded regions.  For comparison, the horizontal dashed lines illustrate no correlation.  Table~\ref{tab:FitsTab} reports information on these fits in the following columns: (1) data plotted; (2) slopes of unweighted data with $1\sigma$ errors; (3) slopes of data weighted by cluster stellar masses; (4) reduced $\chi^{2}$ values of these best fit lines; (5) Spearman's rank coefficient ($\rho$), calculated using the data points from all individual clusters; and (6) statistical significance of $\rho$ (P).  Spearman's $\rho$ can range from $-1$ (inverse correlation) to 0 (no correlation) to $+1$ (direct correlation), and a low P indicates a significant $\rho$ value.  Columns 4-6 are calculated from the unweighted data.  Using bin medians instead of averages does not significantly change the results.

All plots show direct correlations with varying degrees of significance at close to or greater than $2\sigma$.  The plot versus $p_{\Delta}$ exhibits a trend with a significance greater than $3\sigma$, and two of the plots -- SF fraction vs. number of components and $p_{\alpha}$ -- show trends at greater than $2\sigma$ significance.  SF fraction versus $p_{\beta}$ exhibits a trend with very slightly less than $2\sigma$ significance.  When the data are weighted by cluster mass (as explained above), the significance of all the plots increases slightly or remains constant.  Removing the upper limits from the fits in Figure~\ref{fig:SFfracvspFig} retains the direct correlations observed but decreases the significances of the correlations due to increased uncertainty in the fit.

All of these results suggest that a weak direct correlation exists between SF in, and global substructure properties of, clusters.  It is promising that all four measures of substructure exhibit a similar positive correlation with SF fraction.  On average, more substructure appears to be correlated with total cluster SF, in agreement with our analysis in \S\ref{sec:SFvsDistDensSec}.

We note that the SF fractions in the five- and six-component clusters in Figure~\ref{fig:SFvsNumCompsFig} are strikingly lower than the fit would suggest.  Including these in the fit produces slopes consistent with zero.  These points could also indicate a possible maximum number of components at which SF in clusters is greatest.  To test this, however, more clusters with greater than four components than our sample provides must be studied.

\section{Discussion}
\label{sec:DiscussionSec}

We find higher SF fractions in multi-component clusters than in single-component clusters at almost all clustercentric distances and local densities in \S\ref{sec:SFvsDistDensSec}.  This difference is also reflected in the average SF fractions of all single- and multi-component clusters.  We also reproduce the well-established correlation between SF and clustercentric distance, and the inverse correlation between SF and local galaxy number density, in both single- and multi-component clusters.  Furthermore, we find a weak but significant correlation between SF and the amount of global cluster substructure in \S\ref{sec:SFvsClusterSec} in all of our comparisons utilizing statistical test $p$-values and number of components.  These results suggest that the presence of substructure is related to a greater presence of SF in clusters, and the amount of such substructure on average weakly contributes to this relationship.  Furthermore, this weak difference in SF between single- and multi-component clusters is consistent with the fact that little SF is found in clusters at low redshift \citep[e.g.,][]{Popesso2012, Webb2013}, even among merging clusters.

This higher SF fraction is not, however, preferentially found in regions between component centers, as shown in \S\ref{sec:SFvsCompsSec}.  Across our entire population of two- and three-component clusters, we do not find higher fractions of SF in regions between component centers.  Indeed, we find that regions between component centers have a lower average SF fraction than regions elsewhere, consistent with the higher number densities found between component centers.

The trends between amount of SF and substructure may indicate that, on average, the process of cluster mergers actively enhances SF in member galaxies.  However, we propose a different explanation: clusters with substructure are less evolved than unimodal clusters and therefore exhibit SF closer to the field value.  Galaxies in clusters exhibit, on average, lower SF than their counterparts in the field, due to quenching from a variety of gravitational and hydrodynamical mechanisms.  Multi-component clusters are in the process of growing via interactions, so SF in these galaxies is still being quenched and decreasing from the field value.  On the other hand, single-component clusters have had more time after any past interaction to reach a relaxed state in which quenching more strongly affects member galaxies.

These results are in contrast to another study of the relationship between substructure and SF in a large sample of clusters.  \citet[][hereafter A10]{Aguerri2010} studied 88 nearby ($z < 0.1$) clusters using data from the SDSS-DR4, supplemented with the NASA Extragalactic Database (NED).  They found no difference between the fraction of blue galaxies in clusters with and without substructure, measured using the same $\Delta$ test method we employed.  However, important differences exist between the methods in \citetalias{Aguerri2010} and ours: they only included galaxies within one virial radius of their clusters, while we perform measurements out to several virial radii; they defined a cluster as having substructure if $p_{\Delta} \le 0.05$, while we use the number of components from \emph{Mclust}; they defined a galaxy as blue if it has a color $u - r < 2.22$, as in \citet{Strateva2001}, while we use the method explained in \S\ref{sec:DataSec}; and they calculated the mean blue galaxy fraction for their clusters, while we calculate the SF fraction across our entire sample.

As a check on the result in \citetalias{Aguerri2010}, we performed an analysis on our sample using their methods, though instead of re-calculating the values of $p_{\Delta}$ only using galaxies within one virial radius, we used the values given in \citetalias{Einasto2012a}.  We found no significant difference in mean blue galaxy fraction between clusters with and without substructure, which is in agreement with the result from \citetalias{Aguerri2010}.  However, we suggest that our primary analysis may be more sensitive.  Galaxies in subclusters can be affected by cluster environment well beyond the virial radius, as shown, for example, by simulations in \citet{Dolag2009} and by the consistent difference between single- and multi-component clusters out to large clustercentric distances in our Figure~\ref{fig:SFvsDistDensFig}.  Therefore, the analysis in \citetalias{Einasto2012a} is likely more robust for identifying substructure.  Furthermore, including galaxies out to larger radii increases our galaxy sample size by almost a factor of three, improving the statistical precision.

The large scatter in our observed SF fractions suggests that other cluster properties related to subtructure and merging could also have an effect on SF.  For example, SF ignition by mergers in some clusters and truncation in others would, on average, mostly cancel out while still producing the observed scatter.  It has also been shown that clusters at different stages of merging can show varying degrees of affected SF for the same apparent amount of substructure \citep[e.g.,][]{Hwang2009}.  Estimating merger histories of our clusters using the radial infall model \citep{Beers1982}, however, is currently possible only in bimodal clusters while assuming points masses merging with no angular momentum.  The possible merger histories of some clusters with more than two components have been discussed using detailed study of the minor components' gas and galaxy distributions \citep[e.g.,][]{Hwang2009}, but conclusions about these systems remain speculative.  Still, including merger histories of only our bimodal clusters might uncover more information that could be lost in our current analysis.

Despite these complications, our weak correlation between SF and substructure demonstrates the importance of studying this relationship across a large sample of clusters.  To improve future studies of the effects cluster mergers have on SF, we can improve our statistics by employing larger samples of clusters, and we can include additional factors in our analysis such as merger histories, large scale cluster environment \citep{Einasto2012b}, and galaxy color, type, and age.

\acknowledgements

We thank the referee for invaluable comments and suggestions, and the SDSS team, as well as the MPA/JHU and NYU researchers, for the publicly available data releases and VAGCs.  Funding for the SDSS and SDSS-II has been provided by the Alfred P. Sloan Foundation, the Participating Institutions, the National Science Foundation, the U.S. Department of Energy, the National Aeronautics and Space Administration, the Japanese Monbukagakusho, the Max Planck Society, and the Higher Education Funding Council for England. The SDSS Web Site is http://www.sdss.org/.  The SDSS is managed by the Astrophysical Research Consortium for the Participating Institutions. The Participating Institutions are the American Museum of Natural History, Astrophysical Institute Potsdam, University of Basel, University of Cambridge, Case Western Reserve University, University of Chicago, Drexel University, Fermilab, the Institute for Advanced Study, the Japan Participation Group, Johns Hopkins University, the Joint Institute for Nuclear Astrophysics, the Kavli Institute for Particle Astrophysics and Cosmology, the Korean Scientist Group, the Chinese Academy of Sciences (LAMOST), Los Alamos National Laboratory, the Max-Planck-Institute for Astronomy (MPIA), the Max-Planck-Institute for Astrophysics (MPA), New Mexico State University, Ohio State University, University of Pittsburgh, University of Portsmouth, Princeton University, the United States Naval Observatory, and the University of Washington.  S.A.C. thanks The William H. Neukom Institute for Computational Science for their generous support.  M.E. and J.V. also acknowledge support from the Estonian Ministry for Education and Science research project SF0060067s08, and from the European Structural Funds grant for the Centre of Excellence ``Dark Matter in (Astro)particle Physics and Cosmology" TK120.

\bibliographystyle{apj}

\end{document}